\documentstyle[12pt,epsfig]{article}
\let\section=\subsection  \let\subsection=\subsubsection

\def\be{\begin{equation}}
\def\ee{\end{equation}}
\def\bea{\begin{eqnarray}}
\def\eea{\end{eqnarray}}

\begin{document}
\begin{center}
\rightline{MIT--CTP--2842}
\rightline{nucl--th/9903055}
~\\[5mm]
{\large \bf Soliton formation in the Nambu--Jona--Lasinio
model.\footnote{This work is supported in parts by funds provided 
by the FCT, Portugal (Contract PRAXIS/4/4.1/BCC/2753), by the 
U.S. Department of Energy (D.O.E.) under cooperative research
agreement \#DF--FC02--94ER40818 and by the Deutsche
Forschungsgemeinschaft (DFG) under contract We 1254/3-1.}
}\\[5mm]
Jo\~ao da Provid\^encia, Hans Walliser\\[5mm]
{\small \it 
Departamento de F\'{\i}sica, Universidade de Coimbra, 
P-3000 Coimbra, Portugal 
\\[5mm]} 
{and}\\[3mm]
{Herbert Weigel\footnote{Heisenberg--Fellow}\\[3mm]
{\small \it Center for Theoretical Physics\\
Laboratory of Nuclear Science and Department of Physics\\
Massachusetts Institute of Technology\\
Cambridge, Ma 02139} \\[8mm]}
\end{center}

\begin{abstract}\noindent
The soliton formation is considered in the Nambu-Jona-Lasinio model 
with local four quark interaction and various schemes to regularize
the energy contribution of the polarized vacuum. No additional constraints 
are admitted in order to stabilize the soliton. While solitons are
unstable in the proper--time regularized version the three momentum 
cut--off regularization apparently is more appropriate. Using a 
semi--classical approach multi-quark solitons obtained from that 
scheme are discussed. However, no self--consistent non--trivial 
unit baryon number configuration has been found. We also study a 
renormaliz\-able extension of the model. In this case no stable 
multi--quark solitons are obtained within the 
semi--classical approach.
\end{abstract} 

\leftline{PACS 12.39.-x, 14.20.-c}  
\leftline{Keywords: NJL soliton, Wigner--Kirkwood expansion}

\newpage

\section{Introduction}

The strong interaction dynamics of mesons and baryons is described by
quantum chromodynamics (QCD). It is assumed that this theory explains
hadronic phenomena in terms of the quark and gluon degrees of freedom. At 
low energies, comparable with the low lying hadron masses, QCD exhibits 
a non--perturbative behavior. This circumstance renders the analytic 
study of the theory rather difficult. Nevertheless, a qualitative 
description of important aspects of QCD is possible on the basis of 
effective theories which have in common with QCD the symmetries of the 
quark--flavor dynamics, in particular the pattern of spontaneous chiral
symmetry breaking. In this respect the Nambu--Jona--Lasinio (NJL) 
model \cite{njl} has been very successful in the description of the 
meson sector \cite{vw91,hk94}.

It is therefore natural to investigate the soliton solutions of the 
model in order to describe baryons according to the $1/N_C$ expansion. 
In this context the important issue to be addressed concerns the stability 
of these soliton configurations. To be specific we consider the NJL model 
with a local four quark interaction and various regularization prescriptions 
for the contribution of the polarized vacuum to the mass of the soliton.
This is motivated by the common belief that its stability strongly 
depends on the imposed regularization procedures. For example, in refs 
\cite{tueb,bochum} the description of baryons as solitons with three 
valence quarks derived from the NJL model with proper--time regularization  
\cite{s51} is reviewed. For this description to work the scalar
and pseudoscalar fields $\sigma$ and $\mbox{\boldmath $\pi$}$ must
artificially be stabilized by the so--called chiral circle condition
$\sigma^2 + \mbox{\boldmath $\pi$}^2 = \mbox{const.}$. However, it is 
known that without this condition the soliton using 
proper--time regularization is not stable 
\cite{wt92,smgg92}. Here, we wish to define the stable soliton as 
a non--trivial mean-field solution, which is stable by itself once the
regularization prescription has been fixed. In particular we do not
allow for additional constraints which artificially stabilize 
the field configuration but do not directly follow from the model. 
This criterion is apparently not met by the just mentioned soliton.

Another example is the application of the proper--time regularization 
to the imaginary (anomalous) part of the effective action as well 
\cite{swra93}\footnote{This term is conditionally finite and hence
does not require regularization. Its regularization might spoil the
anomaly structure of the model.}. As a consequence the baryon number 
is no longer quantized and may assume non--integer values. The soliton 
can then be stabilized by constraining the regularized baryon number 
e.g. to unity. As this stabilization mechanism originates from a 
particular regularization prescription the question of whether or not
it fulfills our above criterion remains a matter of taste.
Alternatively one might add further interaction terms to the 
model Lagrangian. In that respect the inclusion of a Higgs--type
meson self--interaction, i.e. 
$(\sigma^2+\mbox{\boldmath $\pi$}^2-\mbox{const.})^2$, 
has been shown to render the soliton stable \cite{waw93}.

Not to be misinterpreted, we do not regard these additional 
constraints as unreasonable. The chiral circle condition as suggested 
by Skyrme type models, which have this condition implemented by
definition, may certainly be in accordance with a reasonable description 
of single baryons as solitons. The baryon number constraint to integer 
values is obvious, and the forth--order self--interaction is unavoidable 
in sensibly renormalizable models. An additional forth--order 
self--interaction may also be motivated by higher
multi--quark interactions following from QCD when gluons are integrated
out \cite{w97}. The point, however, is that these additional 
constraints do not directly follow from the NJL model together with 
an adopted regularization procedure.

Recently it has been shown \cite{fprs}, using the Thomas--Fermi 
method, that the NJL model with three dimensional momentum cut--off 
regularization of the Dirac sea, apparently accounts for the existence of 
stable solitons, provided the number of valence quarks is large 
enough. Needless to emphasize that the phenomenology of multi--quark
or multi--baryon systems requires the $\sigma$ and $\mbox{\boldmath $\pi$}$
fields to be treated independently. We compute the  
critical size of this soliton below which it ceases to be stable. 
This size is determined by the interplay between the bulk and 
surface energy densities. For the proper description of the surface 
the lowest order gradient corrections to the Thomas--Fermi method 
in the Wigner--Kirkwood expansion are taken into account.
These investigations are described in section 2 and are completed 
in section 3 with a self-consistent calculation for unit baryon number
imposing the corresponding regularization scheme on the single 
particle Dirac Hamiltonian. 

Finally, in section 4, we briefly discuss a renormalizable extended 
version of the NJL model whose solitons are yet unstable, apparently for
the same reason as those in the proper--time regularized version.
Our conclusions are summarized in section~5.

\section{Thomas--Fermi method and Wigner--Kirkwood expansion}

Our starting point is the semi--bosonized version of the NJL model
where the scalar and pseudoscalar fields $\sigma$ 
and $\mbox{\boldmath $\pi$}$ have been
introduced through a Hubbard--Stratonovich transformation 
to eliminate the original four--fermion interaction
\be\label{njl}
{\cal{L}}_{NJL}=\bar q \left[ i \gamma^\mu \partial_\mu - 
(\sigma + i \gamma_5 \mbox{\boldmath $\tau$}
\cdot\mbox{\boldmath $\pi $})
\right] q - \frac{\sigma^2 + \mbox{\boldmath $\pi$}^2}{2 G}
\, . 
\ee
We do not consider explicit chiral symmetry breaking. In the vacuum 
sector the scalar field is identical to the constituent quark mass $m$. 
In order to prevent the model from becoming a trivial theory of 
non--interacting mesons a regulator has to be retained or alternatively 
the necessary counterterms have to be added to the Lagrangian (\ref{njl}). 
In the subsequent section we will study the three--momentum cut--off in 
more detail because this regulator suppresses the high momentum 
components, which in the proper--time regularization scheme cause 
the soliton to shrink to a point--like singularity \cite{wt92,smgg92}.
For that reason we present the following expressions in that
particular scheme, the generalization to other regularization 
schemes is straightforward. The three--momentum cut--off $\Lambda$ 
is related to the four--fermion coupling $G$ and the constituent quark 
mass $m$ via the gap--equation
\be\label{gap}
\frac{1}{G} = 
4N_C \!\! \int \limits_{|{\bf p}| \leq \Lambda} \!\! \frac{d^3p}{(2\pi)^3} 
\frac{1}{\sqrt{p^2+m^2}}
\, .
\ee
In the chiral limit the pion decay constant obtained from
\be\label{fpi}
f^2_\pi= 
N_C m^2 \!\! \int \limits_{|{\bf p}| \leq \Lambda} \!\! \frac{d^3p}{(2\pi)^3} 
\frac{1}{(p^2+m^2)^{3/2}}
\, ,
\ee
will later be used to determine the value of the constituent mass 
$m$ as a function of the cut--off $\Lambda$.

The semi--classical, relativistic particle and energy densities for
fermions in the background of a scalar field are derived within the 
Wigner--Kirkwood (WK) expansion up to order $\hbar^2$ \cite{wk1,wk2,wk3}
\be\label{densities}
\rho = \rho^{(0)} + \rho^{(2)} \, , \qquad
{\cal{E}} = {\cal{E}}^{(meson)} + {\cal{E}}^{(0)} + {\cal{E}}^{(2)}
\, .
\ee
The leading order (Thomas--Fermi method) was already investigated
in Ref.\cite{fprs} and the next to leading order includes
gradient corrections
\bea\label{expansion}
&&\rho^{(0)} = - 4N_C \!\! \int \limits_{p_F \leq |{\bf p}| \leq \Lambda} 
\!\! \frac{d^3p}{(2\pi)^3} \nonumber \\
&&\rho^{(2)} = \frac{N_C}{6} \!\! 
\int \limits_{p_F \leq |{\bf p}| \leq \Lambda} 
\!\! \frac{d^3p}{(2\pi)^3} \frac{1}{p^4}
   \left[ (1-\frac{3\sigma^2}{p^2}) (\nabla \sigma)^2 - 2 
\sigma \Delta \sigma \right]
\nonumber \\
&&{\cal{E}}^{(meson)} = \frac{\sigma^2}{2 G} \\
&&{\cal{E}}^{(0)} = - 4N_C \!\! \int \limits_{p_F \leq |{\bf p}| \leq \Lambda} 
\!\! \frac{d^3p}{(2\pi)^3}  \sqrt{p^2 + \sigma^2} 
\nonumber \\
&&{\cal{E}}^{(2)} = \frac{N_C}{6} \!\!
\int \limits_{p_F \leq |{\bf p}| \leq \Lambda} \!\! 
\frac{d^3p}{(2\pi)^3}  
   \frac{\sqrt{p^2 + \sigma^2}}{p^4}
   \left[ (1-\frac{3\sigma^2}{p^2}) (\nabla \sigma)^2 - 2 
\sigma \Delta \sigma \right]
\, .\nonumber 
\eea
Note that the scalar field is space--dependent but does not depend 
on the momenta. For the valence contribution the three--momenta were 
integrated from zero to the Fermi momentum $p_F$. To this the  
contribution of the polarized vacuum has been added. It is obtained
from the same integrals, however, with the opposite sign and the upper
bound changed
to the cut--off $\Lambda$. This not only explains the integration
bounds in eq. (\ref{expansion}) but also removes the infra--red
singularities dwelling in the individual contributions to the
gradient terms.

In principle pseudoscalar fields may be introduced in a chirally
symmetric way by the replacements 
$\sigma^2 \to \sigma^2 + \mbox{\boldmath $\pi$}^2$,
$(\nabla \sigma)^2 \to (\nabla \sigma)^2 +
(\nabla \mbox{\boldmath $\pi$})^2$ and
$\sigma \Delta \sigma \to \sigma \Delta \sigma +
\mbox{\boldmath $\pi$}\cdot \Delta \mbox{\boldmath $\pi$}$.
We do not give the explicit expressions because within the
semi--classical approach the pseudoscalar fields are not excited.
The densities (\ref{expansion}) 
\bea\label{soliton}
& \bar \rho =  \rho - \rho^{(vac)} \, , \qquad
& \rho^{(vac)} = - 4N_C \!\! \int \limits_{|{\bf p}| \leq \Lambda} \!\! 
\frac{d^3p}{(2\pi)^3}  
\\
& \bar {\cal{E}} = {\cal{E}} - {\cal{E}}^{(vac)} \, , \qquad
& {\cal{E}}^{(vac)} = \frac{m^2}{2 G}
- 4N_C \!\! \int \limits_{|{\bf p}| \leq \Lambda} \!\! 
\frac{d^3p}{(2\pi)^3}  \sqrt{p^2 + m^2} 
\nonumber
\eea
determine the quark number and the energy measured
relative to the vacuum. 
All further applications follow from the variation of the functional
\be\label{variation}
\delta \int d^3 r ( \bar {\cal{E}} - \lambda \bar \rho) = 0 \, ,
\ee
where the  
Lagrange multiplier $\lambda$ (chemical potential) has been
introduced to fix the quark number $A$.
This constraint should not be confused with the one employed in
\cite{swra93}. Here it is due to the semi--classical treatment
and serves solely to determine the Fermi momentum $p_F$. In a
microscopical calculation as e.g. the one presented in subsection 3.3
no such constraint is required.  

Subsequently we apply this method to hadronic matter and to finite
hadronic systems.

\section{Three dimensional momentum cut--off regularization}

Apparently the three--momentum cut--off regularization for the 
vacuum contributions is a good candidate to yield stable solitons. 
It successfully suppresses the high momentum components contained
in the infinitely high and narrow peak of the scalar field
at the origin which 
arises e.g. in the proper--time regularization scheme \cite{wt92,smgg92}.
In the following subsections we apply the Thomas--Fermi method 
together with its gradient corrections to hadronic matter and to
finite hadronic systems. Finally, in subsection 3.3, we present
a fully self--consistent calculation for unit baryon number.

\subsection{Hadronic matter}

For hadronic matter the gradient terms in (\ref{expansion})
do not contribute, only the leading order in the WK
expansion which corresponds to the Thomas--Fermi method survives.
The variation (\ref{variation}) with
respect to the Fermi momentum $p_F$ and the scalar field
$\sigma$, respectively, leads then to
simple expressions for the particle number 
and the energy discussed already in \cite{fprs}
in connection with the chiral phase transition. 
The binding energy per quark $\bar {\cal{E}} / \bar \rho -m$ 
is plotted in Fig.1 of that reference for various ratios of the 
cut--off in units of the constituent mass.
The maximal binding energy (minima of those curves) is always reached
in the symmetric phase, $\sigma \equiv 0$, at
\bea\label{minimum}
&&\bar {\cal{E}} / \bar \rho -m = p_F -m \\
&&p_F^4 = \frac{3}{2} \left[ \Lambda \sqrt{\Lambda^2+m^2}
(2 \Lambda^2 - m^2) + m^4 \ell n \frac{\Lambda + 
\sqrt{\Lambda^2+m^2} }{m} - 2 \Lambda^4 \right] \, ,  
\nonumber 
\eea
which in units of the free constituent quark mass $m$ is 
the function of $\Lambda / m$ 
\begin{figure}[htbp]
\centerline{\hspace{-3cm}
\epsfig{figure=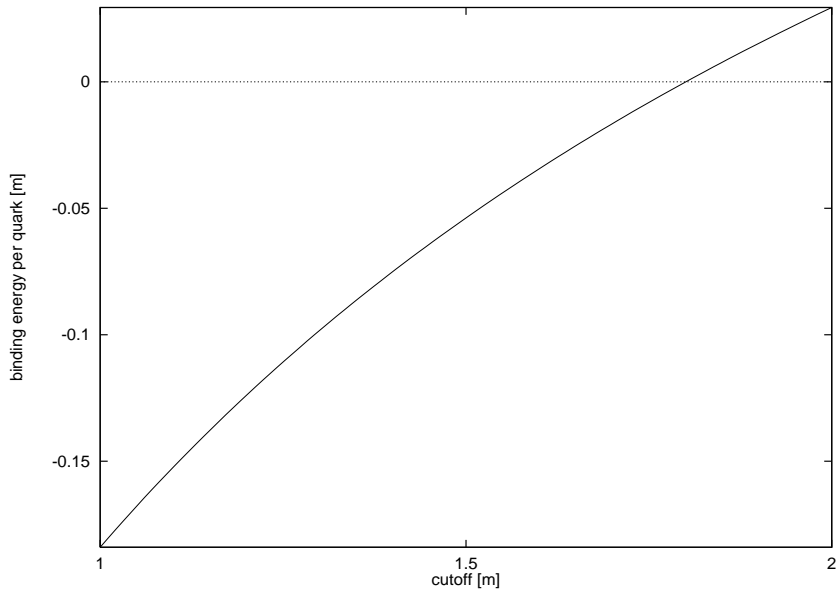,height=8cm,width=8.0cm}}
\protect\caption{Binding energy per quark for hadronic matter
in dependence of the cut--off $\Lambda$. Solitons are obtained
for $\Lambda /m \leq 1.80$.
}
\end{figure}
depicted in our Fig.1. From this figure we notice that hadronic
matter becomes bound for $\Lambda /m \leq 1.80$, for larger
ratios no solitons exist. Since for finite systems the gradient 
terms (cf. next subsection) introduce surface repulsion,
the latter statement is quite general.
With decreasing ratio $\Lambda /m$ we expect to find solitons also
for a smaller number of quarks. For what follows we choose
$\Lambda /m = 1.7, 1.6$ and $1.5$ which corresponds to a NJL
coupling $G=1.60, 1.86$ and $2.18 \, m^2$ according to the 
gap-equation (\ref{gap}). All dimensionalfull quantities will be
expressed in terms of the free constituent mass $m$. This scale
may be fixed by the empirical value of the pion decay constant
$f_\pi \simeq 90$MeV in the chiral limit using (\ref{fpi}).
\begin{table}[htbp]
\begin{center}
\parbox{14.5cm}{\caption{\label{tab_1}
Parameters of the model for various ratios of the cut--off divided
by the constituent mass. The pion decay constant is kept fixed at
$f_\pi=90$MeV. The last two columns give the quark condensate
and  the binding energy per quark in hadronic matter.
}} 
\begin{tabular}{|c|ccc|c|c|}
\hline
$\Lambda / m$ & $m$ [MeV]    & $\Lambda$ [MeV] & $G$ [GeV$^{-2}$] & 
$(-<\bar q q>)^{\frac{1}{3}}$[MeV]   & $\bar {\cal{E}} / \bar \rho -m$ [MeV] \\
\hline
$1.7$      & $348$   & $592$   & $13.2$ & $298$  & $ -5.8$   \\
$1.6$      & $365$   & $583$   & $14.0$ & $297$  & $-12.5$   \\
$1.5$      & $383$   & $575$   & $14.8$ & $296$  & $-20.6$   \\
$\cdots$   &         &         &        &        &           \\
$1.0$      & $553$   & $553$   & $20.2$ & $301$  & $-101.8$   \\
\hline
\end{tabular}
\end{center} 
\end{table}
The corresponding
model parameters and the results for the binding energy per
quark in hadronic matter are listed in Table 1. The last row
refers to an unphysically low value for the cut--off $\Lambda/m = 1.0$
with strong coupling $G=6.17 m^2$ leading to a considerable increase 
of the binding energy per quark. We provided this parameter set for 
later reference in subsection 3.3, where we discuss the three 
quark soliton as obtained in a self-consistent calculation.

In the following subsection we are taking the next to leading terms
in the WK expansion into account in order to study the
soliton formation in finite systems.

\subsection{Finite hadronic systems}

Here we consider the second order terms in the WK expansion 
(\ref{expansion}), whose gradient terms contribute a repulsive 
mesonic kinetic energy corresponding to the surface repulsion 
of the soliton. Since this semi--classical expansion is designed 
for large particle numbers, spin--isospin uncorrelated quark states 
are assumed. Consequently the pseudoscalar field enters
these expressions at least quadratically and the variation
(\ref{variation}) makes this field vanish identically. Note 
that the non--zero value of the scalar field in the vacuum
provides the driving term for $\sigma(r)$. For symmetry
reasons it is a radial function.

The variation (\ref{variation}) with
respect to the Fermi momentum $p_F(r)$, which is now a
radial function as well, leads to
\bea\label{fermi}
&\epsilon_F = \sqrt{p_F^2 + \sigma^2} = \lambda     
\qquad \qquad \qquad \qquad \qquad & r \leq R  
\qquad \mbox{`inside'} \nonumber \\
&(1-3 \sigma^2 / p_F^2) (\nabla \sigma)^2 - 2 \sigma \Delta \sigma 
= 24 p_F^4  \qquad & r > R \qquad \mbox{`outside'} \, .
\eea
The ``radius'' $R$ divides coordinate space into an interior region, 
where the Fermi
energy is fixed by the chemical potential as in the hadronic matter
case, and into an outer region, where in the Thomas--Fermi
approximation $p_F=0$ when the gradients are neglected. The
presence of these gradient terms prevents $p_F$ and
hence the quark density from becoming exactly zero in the outer
region. Instead all these quantities will obtain a smooth exponential
tail.

In the interior region a non--linear differential equation
follows from variation (\ref{variation}) with respect to $\sigma(r)$.
For completeness
we give this equation explicitly  
\bea\label{deq}
\frac{\sigma}{G} &+&
\frac{N_C \sigma}{\pi^2} \left[ \epsilon_F p_F -
\epsilon_\Lambda \Lambda
-\sigma^2 \ell n \frac{p_F + \epsilon_F}{\Lambda + \epsilon_\Lambda}
\right] \nonumber \\
&+& \frac{N_C}{12 \pi^2} \left[ 2 \left( -\frac{\epsilon_F}{p_F} 
+\frac{\epsilon_\Lambda}{\Lambda} (3-\frac{\epsilon_\Lambda^2}{\Lambda^2}) 
- 3\frac{\epsilon_F}{\Lambda}
+ 3\ell n \frac{p_F + \epsilon_F}{\Lambda + \epsilon_\Lambda} \right)
\Delta \sigma  \right. \\
&& \quad + \left. \left(
\frac{\epsilon_\Lambda}{\Lambda} (4-\frac{\epsilon_\Lambda^2}{\Lambda^2}) 
-\frac{\epsilon_F}{p_F} (2+\frac{\epsilon_F^2}{p_F^2})
-2\frac{\sigma^4}{\epsilon_\Lambda \Lambda^3} 
+2\frac{\epsilon_F \sigma^2}{\Lambda^3} \right)
\frac{(\nabla \sigma)^2}{\sigma} \right] = 0 \, ,
\nonumber
\eea
where we have introduced the abbreviation
$\epsilon_\Lambda=\sqrt{\Lambda^2 + \sigma^2}$.
This second order differential equation, subject to appropriate
boundary conditions, is then solved for a given chemical potential
$\lambda=\epsilon_F(R) = \sqrt{p_F^2(R) + \sigma^2(R)}$ related to
a corresponding quark number $A$. In practice this is achieved
by fixing the radius parameter $R$.

In principle, a similar differential equation has to be solved 
in the outside region together with (\ref{fermi}), resulting in 
an additional set of two coupled non--linear second order 
differential equations for the functions $\sigma(r)$ and $p_F(r)$. 
Fortunately for large distances the asymptotical solution is known 
analytically
\bea\label{asymptotical}
&&\sigma(r) \stackrel{r \to \infty}{=}  m - \mbox{const.} 
\cdot e^{-2 m r} /r \, , \nonumber \\
&&\rho (r) - \rho^{(vac)} \stackrel{r \to \infty}{=}
\frac{2N_C}{\pi^2} p_F^3(r) \, , \qquad \qquad
p_F^4(r) \stackrel{r \to \infty}{=} \frac{1}{3} m^3 (m-\sigma(r))
\, , \nonumber \\
&&{\cal{E}} (r) - {\cal{E}}^{(vac)} \stackrel{r \to \infty}{=} 
\frac{2N_C}{\pi^2} m p_F^3(r) \, . 
\eea
The constant appearing in $\sigma(r)$ is determined by the 
boundary conditions at $r=R$.
We assume these solutions to be valid in the entire outer region,
which is justified by the observation that at the radius $R$ the
profile $\sigma(R)$ is already close to its asymptotical value $m$
(e.g. for $\Lambda / m =1.5$ with $A=12$ quarks
we have $\sigma(R)=0.943 \, m$ at $R=3.12 \, m^{-1}$, cf.  
solid curve in Fig.3).
\begin{figure}[htbp]
\centerline{\hspace{-3cm}
\epsfig{figure=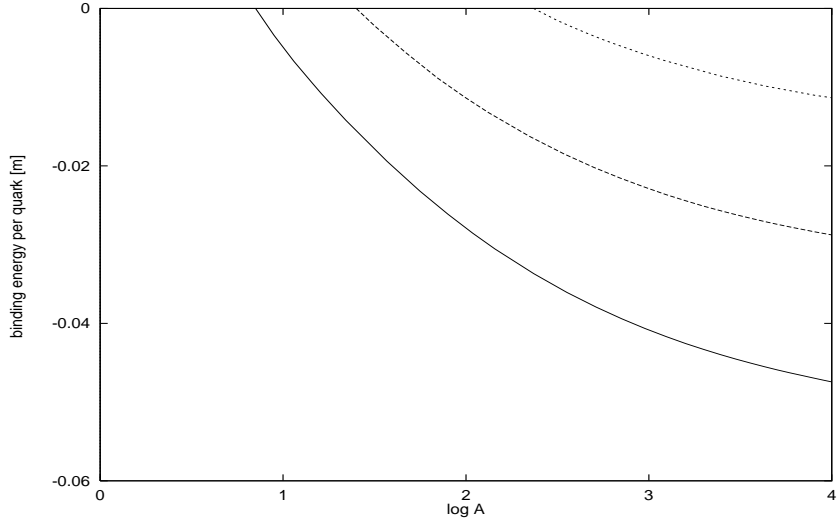,height=7cm,width=8.0cm}}
\protect\caption{Binding energy per quark as a function of
the number of quarks $A$. The solid, dashed and dotted lines
corresponds to different values of the cut--off
$\Lambda /m = 1.5, 1.6$ and $1.7$ respectively. The corresponding
infinite matter binding energies (Table 1) are $-0.054, -0.034$
and $-0.017 \, m$. 
}
\end{figure}
Using a logarithmic scale the resulting binding energy per quark 
$\bar {\cal{E}} / \bar \rho - m$ is plotted in Fig.2 as a function 
of the quark number $A$ for various values of the three--momentum 
cut--off. The binding is always weaker than
that of hadronic matter which is slowly approached with
increasing particle number. The surface effect
increases rapidly with decreasing particle number such that there
exists a critical number of quarks for which the soliton
ceases to exist. 
\begin{figure}[htbp]
\centerline{\hspace{-3cm}
\epsfig{figure=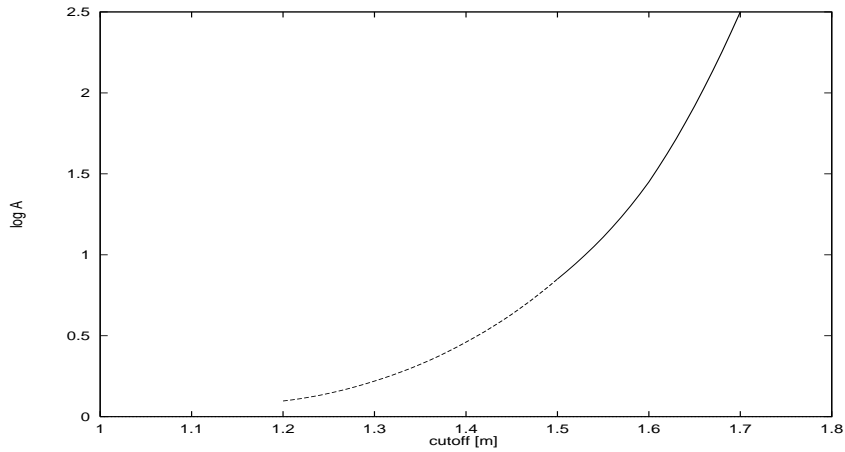,height=6cm,width=8.0cm}}
\protect\caption{Critical number of quarks in dependence of
the cut--off $\Lambda/m$ as obtained in the semi--classical
approximation. Above the critical curve soliton
formation is possible. The dashes indicate that this result
should not be trusted for small quark numbers.
}
\end{figure}

This number sensitively depends on the ratio $\Lambda/m$ which 
controls the coupling constant $G$ and the strength of binding in 
the NJL model with three--momentum cut--off. This dependence is shown 
in more detail in Fig.3 where the quark number is again plotted 
logarithmically. For $\Lambda/m > 1.8$, as we have seen above, quark
matter becomes unbound. With decreasing ratio then also solitons with a
smaller number of quarks become bound, e.g. for $\Lambda/m = 1.5$
we may expect solitons with $A \stackrel{>}{_\sim} 10$ quarks to exist.
However, we should be cautious in trusting our results for too
small quark numbers. Already for $A=12$ with $\Lambda/m=1.5$
(solid curve in Fig.2) the second order WK contribution to particle 
number and energy respectively, amounts to $20\%$ and we may 
expect an error of about $5\%$ from neglecting the higher orders 
in this expansion. In particular, for the case of special interest, 
$A=3$, a hedgehog solution with non--vanishing pseudoscalar field is 
required. This should alter the results considerably. 
\begin{figure}[htbp]
\centerline{\hspace{-3cm}
\epsfig{figure=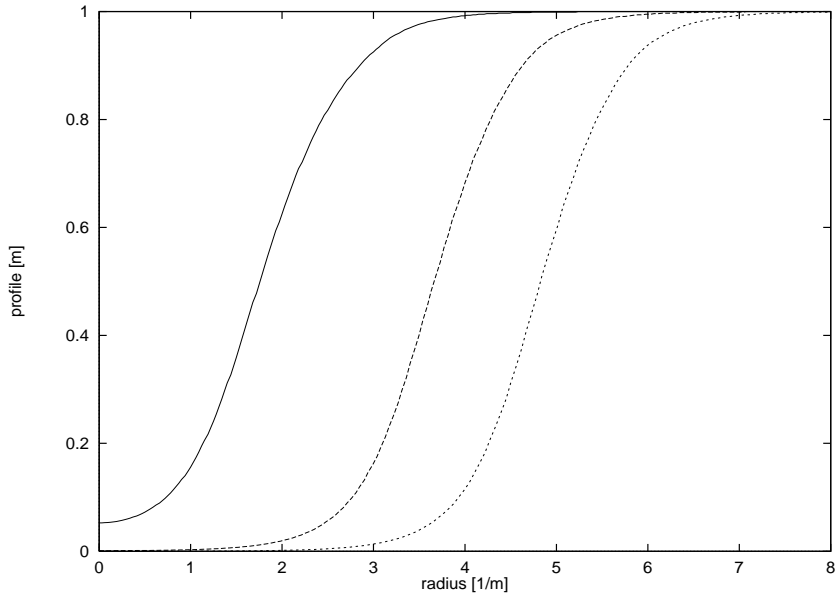,height=8cm,width=8.0cm}}
\protect\caption{Profiles $\sigma$ for
$\Lambda / m =1.5$ as function of the radius in inverse quark masses.
The solid, dashed and dotted lines represent $A=12, 60$ and $120$
quarks.
}
\end{figure}
Finally in Fig.4 we show the calculated profiles for $\Lambda / m =1.5$
for various quark numbers. For large particle numbers 
($A=60, 120$ etc.) the profiles differ only by their radii.
Inside they are essentially zero and outside they approach the
free constituent mass, both regions being smoothly connected 
at the surface of the soliton. For small particle numbers
($A=12$, solid line) the attraction is no longer strong enough
to drive the scalar field into the interior region to 
zero, instead the profile starts in the origin at a finite value.

\subsection{Self--consistent calculation for $A=3$}

Here follows a short description of searching for a self-consistent
configuration with unit baryon number in the three momentum cut--off 
regularization scheme. For this purpose we consider the following 
energy functional
\be
{\cal E}[\sigma,\mbox{\boldmath $\pi$}] = 
\frac{N_C}{2} \epsilon_{\rm val}
\left(1-{\rm sign}(\epsilon_{\rm val})\right)
-\frac{N_C}{2}\sum_\mu \epsilon_\mu(\Lambda)
+\frac{1}{2G}\int d^3r 
\left(\sigma^2+\mbox{\boldmath $\pi$}^2-m^2\right)\, .
\label{efunct}
\ee
In contrast to other covariant regularization schemes 
which regularize the functional trace, which arises when
integrating out the fermions, like
{\it e.g.} proper--time \cite{tueb,bochum}, the three momentum 
cut--off scheme is already applied at an earlier stage,
namely at the level of
the single particle Dirac Hamiltonian which is projected onto the 
corresponding subspace in momentum space. That is, the single 
particle energies $\epsilon_\mu(\Lambda)$ are eigenvalues of
\be
h_\Lambda=
P_\Lambda \left(\mbox{\boldmath $\alpha$}\cdot
\mbox{\boldmath $p$} +\beta 
\left[\sigma + i\gamma_5 \mbox{\boldmath $\tau$}\cdot
\mbox{\boldmath $\pi$}\right]\right) P_\Lambda\, ,
\label{hampro}
\ee
where the projection operator acts on the spinorial 
wave--functions,
\be
P_\Lambda \Psi_\mu = \Psi_\mu
\cases{1 \quad{\rm for} \quad 
\langle\mu|\mbox{\boldmath $p$}^2|\mu\rangle
\le\Lambda^2\cr
0\quad {\rm for} \quad
\langle\mu|\mbox{\boldmath $p$}^2|\mu\rangle > \Lambda^2} \, .
\label{projector}
\ee
Technically this projection is accomplished by diagonalizing
the unprojected Dirac Hamiltonian in a basis consisting of
free Dirac spinors with momentum less than the cut--off $\Lambda$. 
That is, the Fock space for quarks on which the baryon is 
built, is truncated by the cut--off \cite{fprs}.
We will comment on the special role of the valence quark level 
later. The baryon number is defined as the asymmetry in the spectrum 
of $h_\Lambda$. Apparently the just defined regularization yields 
unit baryon number for the functional (\ref{efunct}). 

A self--consistent soliton solution would correspond to 
a local minimum of the energy functional (\ref{efunct}).
In contrast to the multi--quark systems discussed above 
we expect a strong coupling to isospin and hence the pions
for a single baryon. Of course, this driving role of the
pions is a well--known feature of the soliton picture for individual
baryons where the pion fields assume a hedgehog form. We
therefore choose the static {\it ans\"atze}
\be
\sigma(\mbox{\boldmath $r$})=\sigma(r)\qquad {\rm and}\qquad
\mbox{\boldmath $\pi$}(\mbox{\boldmath $r$})
={\hat{\mbox{\boldmath $r$}}} \pi(r)
\label{ansatz}
\ee
which introduces the two radial profile functions $\sigma(r)$ 
and $\pi(r)$. These profile functions obey the self--consistency
conditions
\bea
\sigma(r)&=&-\frac{N_C}{4\pi}G \int d\Omega
\left\{\Psi_{\rm val}^\dagger \beta \Psi_{\rm val}
-\frac{1}{2}\sum_\mu \Psi_\mu^\dagger \beta \Psi_\mu\right\}
\label{eqom1} \\
\pi(r)&=&-\frac{N_C}{4\pi}G \int d\Omega
\left\{\Psi_{\rm val}^\dagger i\beta\gamma_5
\mbox{\boldmath $\tau$}\cdot\hat{\mbox{\boldmath $r$}}\Psi_{\rm val}
-\frac{1}{2}\sum_\mu \Psi_\mu^\dagger i\beta \gamma_5
\mbox{\boldmath $\tau$}\cdot\hat{\mbox{\boldmath $r$}}\Psi_\mu\right\}
\label{eqom2}
\eea
with $\Psi_\mu$ being the eigenfunctions of the projected 
Dirac Hamiltonian (\ref{hampro}). Suitably for the spherical
configuration (\ref{ansatz}) the free basis to diagonalize 
$h_\Lambda$ is discretized by requiring appropriate boundary 
conditions for the quark wave--function at a finite but 
large radius, $D$. Of course, the existence of the 
continuum limit, $D\to\infty$ is crucial. States with momenta 
larger than $\Lambda$ are omitted from that basis. In addition
the basis states are labeled by the conserved grand spin
quantum number with the valence level possessing grand 
spin zero. For more details on these technical aspects 
we refer to the literature \cite{kr1,Al94}. 

The search for a self--consistent solution starts with choosing 
a trial configuration for the profile functions $\sigma(r)$ and
$\pi(r)$. The resulting eigenstates of $h_\Lambda$ are 
subsequently employed to update these profile functions 
according to the equations of motion (\ref{eqom1},\ref{eqom2}).
Iteration of this procedure eventually yields a stable 
configuration. Unfortunately we have been unsuccessful
in finding a non--trivial configuration for the model parameters
of table \ref{tab_1} when taking the valence quark level
to be the lowest eigenstate of the projected Hamiltonian.
This includes the extreme case $\Lambda=m$, which supposedly
yields strong binding (cf. table 1).
This means that during the iteration the meson profiles
tend to the vacuum configuration $\sigma(r)=m$ and $\pi(r)=0$.
It should be noted that much care had to be taken in 
gaining this result as there appeared to be spurious 
finite size effects which faked a non--trivial solution. 
Although the typical extension of a soliton solution is
expected to be of the order of $1{\rm fm}$ we were enforced
to take $D=25{\rm fm}$ or larger to avoid these finite 
size effects. 

On the other hand it should be noted that taking the valence level 
as resulting from diagonalizing the projected Hamiltonian $h_\Lambda$ 
corresponds to also regularizing the valence level which is an 
un--common approach\footnote{In the proper--time scheme, for example,
this would correspond to a non--integer baryon number and also 
destabilize the soliton. In the case of the three--momentum cut--off, 
however, demanding unit baryon number does not prohibit the regularization 
of the valence level.}. We have therefore also studied the case 
where the valence quark contribution to the equations of motion
stems from diagonalizing the un--projected Hamiltonian. In that case 
something peculiar happens. Upon iteration the system tends to approach
the un--stable (peaked) configuration observed in the proper--time 
scheme. This configuration in particular is characterized by a 
negative valence quark eigenenergy $\epsilon_{\rm val}$. However, 
as $\epsilon_{\rm val}$ turns negative, it is considered part of the 
distorted vacuum and {\underline{must}} be taken from the projected 
Hamiltonian, which, as discussed above, does not yield stable solutions. 
Hence in the iterative approach the configuration fluctuates about a point 
where $\epsilon_{\rm val}$ flips sign\footnote{We have also considered
the inclusion of a finite `chemical potential' $-m<\mu<m$ against
which we measure the single quark energies. In that case the 
configuration oscillates about the point $\epsilon_{\rm val}=\mu$.}. 
Apparently that is at best a saddle point solution of the equations 
of motion and cannot be considered a real solution.

\section{Renormalizable extension of the NJL model}

As already discussed in connection with the proper--time 
regularization scheme, the scalar field may develop an infinitely 
high and narrow peak at the origin, which finally destroys the 
soliton. It is intuitively clear that any regularization scheme which 
does not suppress the high momentum components contained in that 
peak will encounter this problem. Here, we want to shed some light 
on how this instability occurs. The Wigner--Kirkwood expansion 
(section 2) proves suitable to give a simple explanation of the 
phenomenon.

For definiteness we consider a non--trivial and renormalizable version
of the NJL model, where the necessary counterterms, namely the mesonic
kinetic energies and quartic mesonic interactions, are added to the 
Lagrangian (\ref{njl}). Here we give only a brief description of the
renormalization procedure with particular emphasis on the soliton
sector, for the details we refer to \cite{coimbra}. For simplicity we
consider the scalar field only, pseudo-scalar fields may be added
straightforwardly 
\be\label{njlraw}
{\cal{L}}=\bar q (i \gamma^\mu \partial_\mu - \sigma ) q
+ \frac{f_0^2}{2} \partial_\mu \sigma\partial^\mu \sigma 
- \frac{1}{2G_0} \sigma^2 - \frac{\lambda_0}{2} \sigma^4
\, . 
\ee
This amounts to a linear sigma model coupled to quarks which is known
to be renormalizable \cite{gl69}. 
According to the renormalization prescription
\cite{coimbra} the bare couplings indicated with subscripts zero
are replaced by the renormalized (finite) parameters $g_\sigma$
and $\mu_\sigma$. The quadratic and logarithmic divergencies
(denoted by $I_{quad}$ and $I_{log}$ in \cite{coimbra}) stemming
from the quark loop are isolated and expressed as momentum 
integrals,
\bea\label{njlren}
{\cal A}&=&-iN_C {\rm Tr}\, {\rm log}
\left(i\gamma^\mu \partial_\mu - \sigma \right)+
\int d^4x {\cal{L}}_{\rm M} \nonumber
\\ {\rm with}\hspace{3cm} && \hspace{10cm}~~~ \nonumber \\
{\cal{L}}_{\rm M}&=&
 \frac{1}{2} \left[ \frac{1}{g_\sigma^2} +     
4iN_C \int \frac{d^4p}{(2\pi)^4} 
\frac{(p^2+\frac{1}{3} m^2)}{(p^2-m^2)^{3}} \right]
\partial_\mu \sigma\partial^\mu \sigma                 
\nonumber \\ && \hspace{0.2cm} +
\left[\frac{\mu_\sigma^2}{4g_\sigma^2} -
4iN_C \int \frac{d^4p}{(2\pi)^4} 
\frac{(p^2-2m^2)}{(p^2-m^2)^2}\right]\sigma^2 
\\                                                  
&& \hspace{0.2cm}
-\frac{1}{2} \left[ \frac{\mu_\sigma^2}{4g_\sigma^2 m^2} + 
4iN_C \int \frac{d^4p}{(2\pi)^4} \frac{1}{(p^2-m^2)^{2}}
\right] \sigma^4 \nonumber                                      
\, . 
\eea
Note that the scalar field used here 
$\sigma = m + g_\sigma \sigma^{\prime}$ is connected to $\sigma^{\prime}$ 
defined in \cite{coimbra} and that the parameters are related
$\mu_\sigma^2 = 4 m^2 ( 1 + N_C g_\sigma^2 / 6\pi^2 )$
such that the Nambu relation for the sigma mass $m_\sigma=2m$ holds. 
In principle the action (\ref{njlren}) is well suited to 
study the soliton in the renormalized extension of the 
NJL model. This is in particular the case because the counterterms,
which are completely fixed in the meson sector, also render finite
the fermion determinant in the soliton background. However, this
is technically quite involved because identical regularization
schemes have to be employed for the functional trace
and the counterterms. In a numerical treatment this is technically 
quite complicated\footnote{{\it Cf.} ref \cite{Fa98} for a suitable 
path to approach this problem.} and beyond the scope of the present 
paper.

Instead we would like to gain some first insight by employing the 
much simpler semi--classical method
described in section 2. It is noticed that the infinities of the extra
terms contained in (\ref{njlren}) just cancel those appearing in
${\cal{E}}^{(0)}$ and ${\cal{E}}^{(2)}$ (eq.(\ref{expansion})) when
$\Lambda \to \infty$ ($\rho^{(0)}$ and $\rho^{(2)}$ are finite in
that limit). This is not obvious, apparently the WK expansion
reproduces all singularities of the exact model correctly.
Stated otherwise, the WK expansion allows us to carry out the 
renormalization program analytically and avoiding a {\it numerical}
regularization of the fermion determinant.

As the resulting action is a perfectly finite functional of the
form (\ref{variation}) we can proceed analogously to 
subsection 3.1. We find that quark matter (symmetric phase with
$\sigma=0$) is in principle bound for $g^2_\sigma>24\pi^2/N_C$.
For finite systems the variation in the interior of the soliton
leads to a non--linear differential equation similar to (\ref{deq}).
However, it turns out that in contrast to (\ref{deq}) this differential
equation possesses no stable solutions. Using a relaxation method for
the solution of the non--linear differential equation with appropriate
boundary conditions, the development of the 
infinitely high and narrow peak in the scalar profile at the origin
is observed quite similar as reported in the proper--time
regularization scheme \cite{wt92,smgg92}.
The kinetic part of the functional (\ref{variation}) inside the 
soliton ($\lambda=\epsilon_F =\sqrt{p_F^2 + \sigma^2}$) as it 
follows from (\ref{expansion}),
\be\label{renorm}
\frac{1}{2} \left[ \frac{1}{g^2_\sigma} + 
\frac{N_C}{6\pi^2} (
\frac{\epsilon_F}{p_F} - 3 \ell n \frac{p_F + \epsilon_F}{m} )
\right] (\nabla \sigma)^2  
\, , 
\ee
explains the cause of this result. Close to the center of the soliton, 
where $p_F$ increases towards $\epsilon_F \stackrel{<}{_\sim} m$ 
($\sigma \to 0$), the second term changes sign and eventually 
exceeds the constant and positive contribution from the first term.
As a consequence
the total kinetic energy becomes negative. Exactly this happened in all
considered cases quite independently from the size of the soliton. 
Therefore, within the WK expansion, it is obviously the kinetic energy 
being no longer positive definite which causes the instability of 
the soliton. Of course, this is equivalent to the observation
of an infinitely narrow and high peak in the scalar profile.


\section{Conclusions}

In this short note we have reported on the investigation of the 
soliton formation in the NJL model with local four quark interaction 
and various regularization procedures. Our interest was focused on
solitons in the absence of additional constraints, which are 
commonly imposed in order to achieve stabilization. 

According to the investigations presented here the following 
picture emerges. Whether or not stable solitons may be found 
in the NJL model depends on the chosen regularization scheme.
First of all there
are the more sophisticated regularization schemes as e.g. Schwinger's
proper--time regularization, which do not limit the high
momentum components contained in the Dirac sea.
Solitons in these schemes are unstable because the scalar field
develops an infinitely high and narrow peak at the origin.
Using the WK expansion we were able to show that within a 
renormalizable extension of the NJL model the instability is related 
to the mesonic kinetic energy density not being positive definite. 

Apparently the three--momentum cut--off regularization scheme 
circumvents the above addressed problem by successfully suppressing
the high momentum components contained in the sharp peak
of the scalar field. This regularization scheme is therefore
particularly appropriate to study the soliton formation in the 
unconstrained NJL model. In detail our findings are:

\begin{itemize}

\item Hadronic matter is bound for $\Lambda / m \leq 1.80$.
Since finite systems add surface repulsion no solitons exist 
for larger ratios. 

\item In order to describe finite size effects the lowest
order gradient corrections to the Thomas--Fermi method in the
WK expansion have been considered. We find that the
binding energy per quark is continuously weakened with
decreasing particle number until the soliton ceases to
exist. This behavior, caused by the repulsive
kinetic terms active in the surface of the soliton, 
reveals a peculiarity of the NJL model, which
consequently prefers the formation of large clusters of hadronic
matter. 

\item The critical size, or equivalently the critical 
quark number of the soliton obviously depends sensitively on 
the ratio $\Lambda / m$. We have computed the limiting function
above which soliton formation takes place. Because of the 
semi--classical approximation employed, this result should not be 
trusted for small quark numbers.

\item An extensive search for a 
{\underline{self-consistent}} configuration with unit
baryon number was performed. No solitons were found, not even
with an unphysically low cut--off $\Lambda /m=1$ which should
provide optimal attraction. In all cases the iterative procedure ran
into the trivial configuration. Allowing the distinct valence 
quark not to undergo regularization, the sharp peak in the 
scalar field reappeared.
\end{itemize}

Concludingly we have to state that although we do find solitons in 
the three--momentum regularized NJL model without further constraints,
the results are unsatisfactory in various respects, in particular
of course for the unit baryon number case. It seems that the NJL model 
together with an adopted regularization procedure alone cannot account
for a reliable description of individual baryons as solitons.
However, this may not necessarily be bad news for model builders.
As discussed, the stabilizing extensions of the model can well be 
motivated by phenomenological considerations.

Finally, we would like to further add a few comments concerning the
stability. Within the semi--classical approach the multi--quark solitons
discussed in section 3 are stable against variations of the scalar 
field. Independently we checked that the infinite system is at least
locally stable also when spatially dependent pseudo--scalar fields are
allowed. However, for finite systems, in particular of course for
very small particle numbers, we may not exclude that in a
self-consistent calculation pseudo--scalar fields leading to
instabilities are excited. For the three--quark system we did not find
a non--trivial solution. We may conjecture that the quartic meson
self--interaction inherent in the NJL model is too weak in order
to stabilize this soliton.

On the other hand, with the chiral circle condition imposed, the
proper-time regularized NJL soliton seems to be stable.
Fluctuations in the lowest channels (grand spin) were searched
and no instabilities detected \cite{We95}. It is natural to assume
that this remains still valid when the chiral circle condition
is softened and replaced by an additional quartic meson self--
interaction. Thus, in both cases the hedgehog presumably represents
the stable minimum configuration.

For other versions of the model as e.g. its non-local extensions
\cite{ft92,gbr98} the latter statement is not at all obvious. Although
there the hedgehog solution is reported to be stable with respect
to monopole deformations it is not excluded that this soliton
leaks through other normal modes.

\subsection*{Acknowledgments}
H.Walliser gratefully acknowledges the hospitality extended to him
at the physical department of the university of Coimbra.

\end{document}